\documentclass[12pt]{book}

\usepackage[dvips]{graphicx,color}
\usepackage{makeidx,tsukuba}

\makeauthorindex
\makeindex

\begin{document}

\BookTitle{\itshape The 28th International Cosmic Ray Conference}
\CopyRight{\copyright 2003 by Universal Academy Press, Inc.}
\pagenumbering{arabic}

\chapter{
The Anisotropy Search Program For the Pierre Auger Observatory}

\author{
Roger Clay,$^1$ for the Pierre Auger Collaboration \\
{\it (1) Department of Physics, University of Adelaide, S. Australia 5005\\
}}

\section*{Abstract}
The Pierre Auger Southern Observatory in Argentina has begun
taking data as it is being developed up to a final enclosed area
of 3000 square kilometres.  A key aspect of the project is to
provide information on the origin of the highest energy cosmic
rays through understanding the arrival directions of those
particles.  To avoid claims of a spurious anisotropy detection
because trials have not been properly accounted, the Auger
Collaboration has agreed '{\it a priori}' to the analysis prescription
presented here.  It specifies the following:

1. The accumulation time for the (future) data set to be analyzed.

2. The anisotropy 'targets', each assigned a chance probability level.

3. The analysis procedure for each trial.

A positive result will be claimed for any target search only if
its chance probability is less than its assigned level.  The
levels are chosen so that the total chance probability for one or
more positive results is 0.001.

Exploratory searches beyond this prescription will be encouraged,
but the Auger Collaboration will not assign any confidence level
to anisotropies that may be discovered that way.  Any such
discovery would identify a good target for a prescription to be
used with a subsequent Auger data set.

\section{Introduction: The Need for a Prescription}

A major aim of the Pierre Auger Collaboration is the identification of deviations 
from
isotropy (the anisotropy) in the flux of cosmic rays.
This aim is based on the expectation that rigidity dependent propagation will
begin to allow a form of directional astronomy with cosmic rays at the 
highest energies.  In past studies, the inevitable statistical noise in the 
data, and subtle time variations in the array stability, have caused some 
claims to be made for 
observed anisotropies which have not been confirmed.  This has partially 
resulted from an inadequate knowledge of the number of trials involved in the 
analysis.  The Auger project aims to avoid the latter uncertainties by an 
{\it a priori} definition of the analysis process  
so that the chance probability of any positive detection 
can be rigorously evaluated.  
We will refer to this as
an {\it a priori} analysis prescription.

A prescription for an anisotropy search specifies the search procedure 
{\it a priori} 
so that a positive effect can be published with confidence in its statistical 
significance. It protects the limited number of important 
searches by documenting ahead of time the search that will be performed on 
a specified dataset.

This does not preclude a thorough exploration of the data for expected or 
unexpected anisotropies but such an exploration would not have immediate discovery 
potential.  It would, however, make a case for the result of that exploration 
to be included in considering a prescription for a future dataset.  We thus 
envisage that new Auger datasets will regularly become available as the array 
continues to operate and a specific and unique prescription will be 
determined {\it a priori} for each of those datasets.

\section{The Prescription}

\subsection{Data Set Characterization}
The Auger array is currently under development and is progressively 
increasing in collecting area.  The early anisotropy data 
will be more complicated to analyse than those from a later period in 
which the array is more stable. However, we wish to test the prescription 
technique and its asociated discipline as early as possible.  We propose 
defining the start
of our first dataset for analysis on 8 August 2003, the end of the ICRC, and 
the end when the array on 16 May 2004, ensuring that the second dataset is
timed to fit with a much more stable period.  The 
event 
directions and energies used will be those produced by the official Auger
reconstruction program in use at the time of the end of the dataset.
We anticipate that this first dataset will contain  about 10$^4$ events with
energies above 1 EeV and about 300 events above 10 EeV.  In the analysis of
this first anisotropy data, we will use events with energies greater than 
1 EeV.

\subsection{The Targets}
{\bf We propose that a publishable positive anisotropy result requires a 
chance probability (after accounting for statistical trials) that is less 
than 0.001.}  We thus propose to prioritize search targets by 
partitioning that overall 0.001 probability.

With our defined dataset, this prescription covers searches either over an 
area specified by a radial distance around a potential source direction, or 
in a specific (point source) direction.  Our prescription is:

{\it Our anisotropy search has a positive publishable result if one or
more of the following is true:

* An excess from the galactic center (RA 17h 42m, dec -29.0$^{\circ}$) 
has a chance probability 
less than .00035 using all data. (15$^{\circ}$ radius)

* An excess from the galactic center has a point source chance probability 
less than .00025 using data with LogE $<$ 18.5. 

* An excess from the AGASA/SUGAR location (b=$0^{\circ}$,l=$7^{\circ}$) 
has a point source 
chance probability less than .00025 using data with LogE $<$ 18.5.

* An excess from NGC0253 (RA 00h 46m, dec -25.3$^{\circ}$) has a chance 
probability less than .00005 
using data with LogE $>$ 19.5. ($5^{\circ}$ radius)

* An excess from NGC3256 (RA 10h 28m, dec -44.2$^{\circ}$) has a chance 
probability 
less than .00005 using data with LogE $>$ 19.5. ($5^{\circ}$ radius)  

* An excess from the center of Cen A (RA 13h 25m, dec -43.0$^{\circ}$) 
has a chance 
probability less than .00005 using data with LogE $>$ 19.5. 
($5^{\circ}$ radius)}
 The
AGASA experiment (Hayashida {\it et al.} 1999), with support from SUGAR data
(Bellido {\it et al.} 2001), has identified a possible
excess from either the galactic centre direction or a close by direction 
within a limited energy range.
That direction has always been an 
{\it a priori} key source for the Auger Observatory and the first prescription 
point reflects this.  The
angular range specified covers the 'SUGAR direction' and also is intended to
cover the expected source extension following propagation in the galactic 
magnetic field at characteristic Auger energies (Clay 2001).  The second and third points relate to those 
observations.  NGC253 is a nearby (distance estimate 2.5Mpc) starburst galaxy 
which has been shown to emit a substantial flux of VHE gamma-ray photons
(Itoh {\it et al.} 2002), NGC3256 is a nearby (33Mpc, Moran {\it et al.} 1999) 
merging multiple galaxy system experiencing highly luminous starburst, and
Cen A is our nearest active galaxy at a distance estimated as 3.4Mpc (Bird and
Clay 1990).  The latter sources are outside the 
galactic plane and the angular range is intended to approximate to a point
spread function for the (largely unknown) halo and intergalactic magnetic 
fields.

\subsection{Evaluation Method}

   The final part of the analysis prescription is to specify how
each of the above probabilities is to be evaluated.  This initial
data set will be small compared to data sets that will accumulate
with the full Auger Observatory, so we cannot expect to
have sensitivity to the small anisotropies that will eventually
be measurable.  Also, this first data set will be taken while the
array is growing and changing shape.  The analysis techniques
prescribed here may be inappropriate for the better data sets of
the future.

An ensemble of simulation data sets will be used to evaluate
the probability that the number of arrival directions in a target
solid angle would be as great or greater than the observed number.
The simulation data sets are constructed from the actual data set
by a 2-dimensional scrambling method.  The method is
therefore testing the null hypothesis that an excess from a
certain region occurred as a statistical fluctuation of isotropy.

    The scrambling method is not perfect.  A large excess from
one part of the sky will distort the sidereal time distribution
that would pertain for isotropic cosmic rays.  The method is
conservative, however, in the sense that the chance probabilities
may be slightly overestimated, making the results appear less
significant than they might actually be.

    The method prescribed here is meant to be explicit.  There is
not space to justify each part of it.  More complicated
procedures might yield more reliable probability evaluations, but
simplicity has been a goal.  Here are the explicit rules:

* Use zenith angles in the range 0-70 degrees.

* Computing on-source number of arrival directions:
  (1) For the simple solid angle targets, an arrival direction is
  in the target if it differs from the target's central
  direction by less than the specified radius.
  (2) For a point source, use a 3-degree Gaussian. 

* Make 5 bins of zenith angles uniformly in cos(theta) between
  cos(theta)=1.0 and cos(theta)=0.5.

* To produce a simulation data set of N directions, sample N
  arrival directions from the real data set.  For each one, keep
  the zenith angle, but sample a new sidereal time from events in
  the same zenith angle bin, and sample a new azimuth also from
  events in that zenith angle bin.  

* If the assigned significance probability is P, do n=1000/P
  simulations in determining the chance probability of the
  observed excess.  This gives an expected number of simulation
  data sets showing a greater positive effect than the real data
  (at the assigned probability) as nP=1000.  The statistical
  uncertainty in the obtained probability would then be roughly
  3\%, i.e. sqrt(1000)/1000.

* These rules can be coded into evaluation programs of different
  languages and in different ways.  Alternative analysis programs
  will be checked against each other using artificial data sets.

We have defined here a specific prescription for the directional analysis of 
the first 
pre-defined Auger data set.  When that analysis is complete, a new 
prescription will be defined for the next data set and so on as the experiment 
continues to collect new data.

\section{References}

\noindent
J.A. Bellido {\it et al.} Astroparticle Physics 15, 167 (2001)

\noindent
Bird, D.J. and Clay, R.W., Proc. Astron. Soc. Aust. 8, 266, (1990)

\noindent
G.L. Cassiday {\it et al.} Nucl. Phys. B (Proc. Suppl.) 14A, 291 (1990)

\noindent
R.W. Clay, Pub. Astron. Soc. Aust. 18, No 2, 148, (2001) 

\noindent
N. Hayashida {\it et al.} Astroparticle Physics 10, 303 (1999)

\noindent
C. Itoh {\it et al.} Astronomy and Astrophysics 396, L1 (2002) 

\noindent
E.C. Moran {\it et al.} Ap.J. 526, 649, (1999)

\endofpaper
\end{document}